\documentclass[reprint,aps, prl, balancelastpage,letterpaper,
floatfix,superscriptaddress]{revtex4-2}
\usepackage[section]{placeins}
\usepackage{amsmath}
\usepackage{amssymb}
\usepackage{amsthm}
\usepackage{bm}
\usepackage{upgreek}
\usepackage{hyperref}
\usepackage{xcolor}
\usepackage{hyperref}
\usepackage{mathptmx}
\usepackage[scaled=0.86]{helvet}
\usepackage[T1]{fontenc}
\usepackage{mhequ}
\let\rho=\varrho

\def\fref#1{Fig.~\ref{#1}}

\definecolor{YKB}{rgb}{0.00,0.20,0.75}
\definecolor{mygreen}{rgb}{0.00,0.65,0.05}
\hypersetup{
  colorlinks=true,
  linkcolor= YKB, 
  linkbordercolor=black, 
  citecolor= YKB,
  urlcolor = YKB
  }
\usepackage{graphicx,epsfig,chemarr}
\usepackage{times}
\def\bm#1{\boldsymbol{#1}}
\usepackage{eucal}
\usepackage{amsmath,amssymb}
\usepackage{relsize}
\usepackage{physics}
\usepackage{comment}
\usepackage{siunitx}
\def\half{{\textstyle\frac{1}{2}}}

\def\ie{{\it i.e.},~}

\def\nins#1{\noindent\emph{#1.}~---~}

\def\eref#1{Eq.~(\ref{#1})}
\def\integer{\mathbb{Z}}
\def\real{\mathbb{R}}

\def\a{{\bf a}}
\def\b{{\bf b}}
\def\c{{\bf c}}

\def\n{{\bf n}}

\def\r{{\bf r}}
\def\w{{\bf w}}

\def\B{{\bf B}}
\def\H{{\bf H}}

\def\R{{\bf R}}

\def\U{{\bf U}}
\def\W{{\bf W}}
\def\Lb{{\bf L}}

\def\SO{\mathrm{SO}}
\def\SU{\mathrm{SU}}
\def\One{\mathbf{1}}
\def\Wlam{\W(\lambda)}
\def\thetasing{\theta_\mathrm{single}}
\def\thetadoub{\theta_\mathrm{double}}
\def\sig{\boldsymbol{\sigma}}
\def\vsig{\pmb{\sigma}}
\def\N{{N}}
\def\L{\ell}
\def\cov{{v}}
\def\tfrac#1#2{\textstyle{\frac{#1}{#2}}}

\newcommand{\ltfrac}[2]{\mbox{\large$\frac{#1}{#2}$}}
\let\theta=\vartheta
\begin{document}

\title{Walks in Rotation Spaces Return Home when Doubled and Scaled}

\author{Jean-Pierre Eckmann}
\affiliation{D\'epartement de Physique Th\'eorique et Section de
  Math\'ematiques, Universit\'e de Gen\`eve, Geneva, Switzerland}

\author{Tsvi Tlusty}%
\affiliation{Department of Physics, Ulsan National Institute of Science and Technology, Ulsan, 44919, Republic of Korea}
\email{jean-pierre.eckmann@unige.ch}
\email{tsvitlusty@gmail.com}

\date{March 7, 2025}

\begin{abstract}
  The dynamics of numerous physical systems, such as spins and qubits, can be described as a series of rotation operations, \ie walks in the manifold of the rotation group. A basic question with practical applications is how likely and under what conditions such walks return to the origin (the identity rotation), which means that the physical system returns to its initial state. In three dimensions, we show that almost every walk in $\SO(3)$ or $\SU(2)$, even a very complicated one, will preferentially return to the origin simply by traversing the walk \emph{twice} in a row and uniformly scaling all rotation angles. We explain why traversing the walk only once almost never suffices to return, and comment on the problem in higher dimensions. 
\end{abstract}

\maketitle

\nins{Introduction}Rotations are omnipresent in physics, governing the dynamics of dipoles, spins, qubits, Lorentz transformations, and numerous other systems. 
A prototypical example is a spin evolving in a time-dependent magnetic field $\B(t)$~\cite{pauli1927}. The motion of a spin state $\uppsi(t)$ is a succession of infinitesimal rotations, each around the instantaneous direction of $\B(t)$
and with an angular velocity (Larmor's frequency) proportional to its magnitude~\cite{marsden2013}. 
Consider now a general time-dependent field $\B(t)$ of duration $T$. The pulse $\B(t)$ may be extremely convoluted, reflecting, for example, many-body stochastic dynamics or passage through complex material. 

A natural question then arises: 
Can one make the field $\B(t)$ return the system
to its original state at the end of the pulse, $\uppsi(T)= \uppsi(0)$? ---  
The chief point of this paper is that this is almost always possible,
no matter how complicated the function $\B(t)$ is (what ``almost always'' means will be explained shortly). 
All that is needed is to apply the pulse sequence $\B(t)$ \emph{twice} or more in a row,
after scaling all rotation angles by a well-chosen factor $\lambda$.
We show that this can be achieved either by uniformly tuning the field's magnitude, $\B(t) \to \lambda \B(t)$ or by uniformly stretching/compressing time, $\B(t) \to  \B(\lambda t)$.

In the special case where $\B(t)$ is confined to a plane, the motion of the spin can be mapped to the rolling motion of a solid body on the plane, along a path specified by $\B(t)$~\cite{rojo2010,segerman2021}. Recently, we designed and constructed solid bodies, dubbed ``trajectoids,'' that can roll indefinitely along general periodic paths~\cite{sobolev2023,eckmann2024}. Based on numerical simulations, we \emph{conjectured} that such bodies exist for almost any path, so that the body returns to its original orientation after two or more periods~\cite{matsumoto2023}. 

This ``two-period'' conjecture, alas, lacked any proof and was limited
to planar fields $\B(t)$. Even worse, it lacked an intuitive
understanding of its geometric origin. In this paper, we \emph{prove} the conjecture for general fields
$\B(t)$ and also explain it as a simple outcome of the distribution of random rotation matrices 
in $\SO(3)$ (and random unitary matrices in $\SU(2)$).
We summarize by discussing the problem in higher dimensions. 

\begin{figure}[htb!]
  \centering\includegraphics[width=0.78\columnwidth]{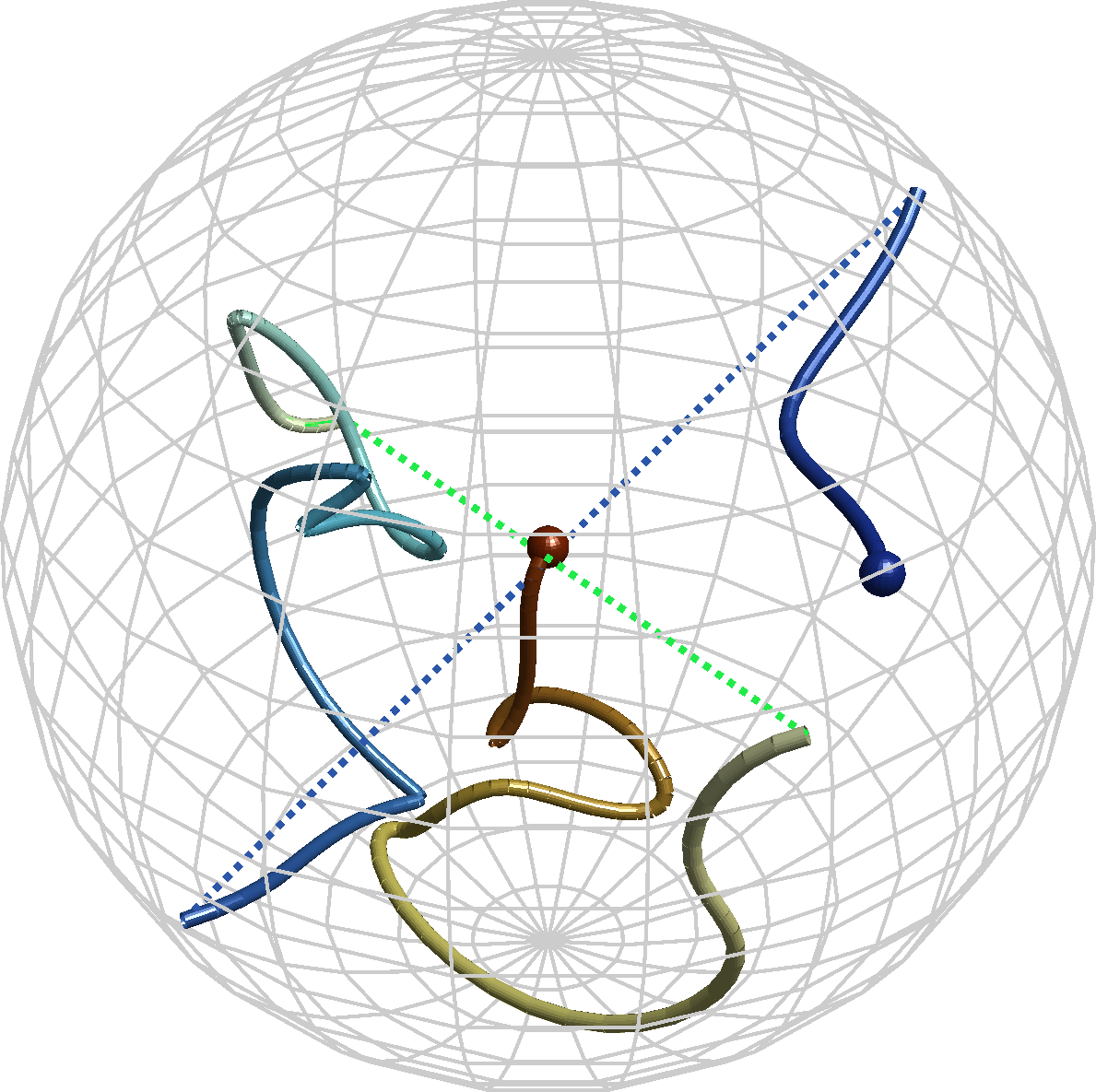}
  \caption{A random walk on $\SO(3)$ shown as a trajectory in a ball of radius $\pi$, where a rotation $\R(\n,\omega)$ is mapped to the point $\r = \n\,\omega$ and antipodal points are identified, $\n \pi = -\n \pi$ (the real projective space $\mathbb{RP}^3$). The walk traverses from the center (small red sphere) to the blue end. Crossing antipodal points is indicated by dotted lines.} \label{fig:sample1}
\end{figure}

\nins{The problem}
A $d$-dimensional rotation, $\R \in \SO(d)$, is specified by $\binom{d}{2}$ parameters, and products of rotations can therefore be seen as walks in a $\binom{d}{2}$-dimensional manifold~\cite{furstenberg1971,varju2013}. Thus, in 3D, a rotation can be specified by $\binom{3}{2}=3$ parameters: for example, a rotation axis $\n$ (a unit vector determined by two angles) and an angle of rotation around it, $\omega \in [0,\pi]$. One can then conveniently visualize 3D rotations as points inside a ball of radius $\pi$ with antipodal points identified (\fref{fig:sample1}). This ball is the real projective space $\mathbb{RP}^3$, which is homeomorphic to $\SO(3)$ by mapping each rotation $\R(\n,\omega)$ to the point $\r = \n \omega$ (see~\cite{altmann2005}[Ch.~10]). Thus, the identity rotation, $\R(\n,0) =\One$, is the ball's center, and \SI{180}{\degree}-rotations compose its surface, where $\R(-\n,\pi) = \R(\n,\pi)$. 

A general product of $\N$ rotations then defines a ``walk'' 
in this ball, $\W =\prod_{j=1}^\N \R_j$ (the time-ordered product
$\R_\N\cdot\ldots\cdot\R_2\cdot\R_1$, Fig.~\ref{fig:sample1}).
Each rotation is specified by its axis and the angle of rotation, $\R_j \equiv \R(\n_j,\omega_j) = \exp \textstyle{[} \omega_j (\n_j \vdot\pmb{L}) \textstyle{]}$, where $\pmb{L} = (\Lb_x,\Lb_y,\Lb_z)$ are the three generators of the group $\SO(3)$.
For $\lambda > 0$, we consider the \emph{scaled} rotations 
$[\R(\n,\omega)]^\lambda =\R(\n,\lambda\omega) = \exp\textstyle{[} \lambda\omega(\n \vdot\pmb{L}) \textstyle{]}$.
To see how the walk $\W$ corresponds to the field pulse $\B(t)$, consider a spin in a magnetic field evolving with a Hamiltonian 
$\H = -\gamma\B(t) \vdot \pmb{S}$, where $\gamma$ is the gyromagnetic ratio
and the spin operator is $\pmb{S} = i \hbar\pmb{L}$. 
It follows that the time evolution operator of the spin is a rotation,
$\exp\textstyle{[}-\frac{i}{\hbar} \H(t) \delta t \textstyle{]}= \exp\textstyle{[}\omega (\n\vdot \pmb{L}) \textstyle{]}$, by an angle $\omega = \gamma \abs{\B(t)} \delta t$ around an axis $\n = \hat{\B}(t)$.
Thus, taking powers of rotations $\R \to \R^\lambda$ corresponds to uniform scaling of the field, $\B(t) \to \lambda \B(t)$, or the time, $\B(t) \to  \B(\lambda t)$.

We ask now: Is there a constant $\lambda > 0$ for which the product of
the powers $\R_i^\lambda$ becomes the unit rotation, $\Wlam
=\prod_{j=1}^\N \R_{j}^\lambda = \One$? 
The transformation $\W \to 
\Wlam$ amounts to the uniform scaling of the individual
rotation angles (or equivalently, the field $\B(t)$).
So, in other words, we ask: What scaling of the angles, $\omega \to \lambda\omega$, will make the transformed walk $\Wlam$ return to the origin, $\Wlam=\One$? 
(in the nontrivial case of  $\N>1$).
In the following, we prove and explain why this problem is generally impossible to solve if the walk is traversed only once. But in striking contrast, we show that the same problem is almost always solvable if the walk is repeated more than once, $[\Wlam]^m=\One$ for an integer $m \ge 2$. 

As we shall see, the underlying geometric reason is that the identity element of $\SO(3)$ is a single point whose codimension is $3$, and is therefore almost impossible to hit by varying a single parameter $\Wlam$. In contrast, there is a whole 2D manifold of roots of identity, with co-dimension one, making it likely that $\Wlam$ will hit one of these roots.

\nins{Probabilistic argument: the ensemble of random rotations}
Before formally proving this result, we explain its origin through an intuitive probabilistic argument. 
To see this, note that, naively, one may try to solve this question by exploring numerous values of $\lambda$ and test whether the scaled walk returns to (or at least approaches) the origin, $\Wlam \to \One$. It turns out that randomly drawing the scaling factor $\lambda$ amounts to randomly drawing rotations, and therefore we need first to explain what ``random'' rotation means. 

A hallmark of randomly drawn rotations is that they evenly distribute points on the unit sphere $S^2$. That is, an ensemble of such rotations maps any given point, say, the north or south pole, to all other points with equal probability. This property defines the invariant Haar measure $\dd {\mu}$ for $\SO(3)$. In the axis-angle representation that we use, $(\n,\omega) \in S^2 \times [0,\pi] \mapsto \R(\n,\omega) \in  \SO(3)$, the Haar measure is uniform with respect to the direction of the axis $\n$, but is biased towards larger rotation angles $\omega$~\cite{miles1965,rummler2002,yershova2010},
$\dd \mu(\n,\omega) =  1/(4\pi^2)(1-\cos\omega) \dd{\omega} \dd{\n}$
(where $\dd \n $ is the uniform measure on $S^2$). This bias in $\omega$ follows from the invariance of the Haar measure to rotations. 

\begin{figure}[htb!]
  \centering\includegraphics[width=0.9\columnwidth]{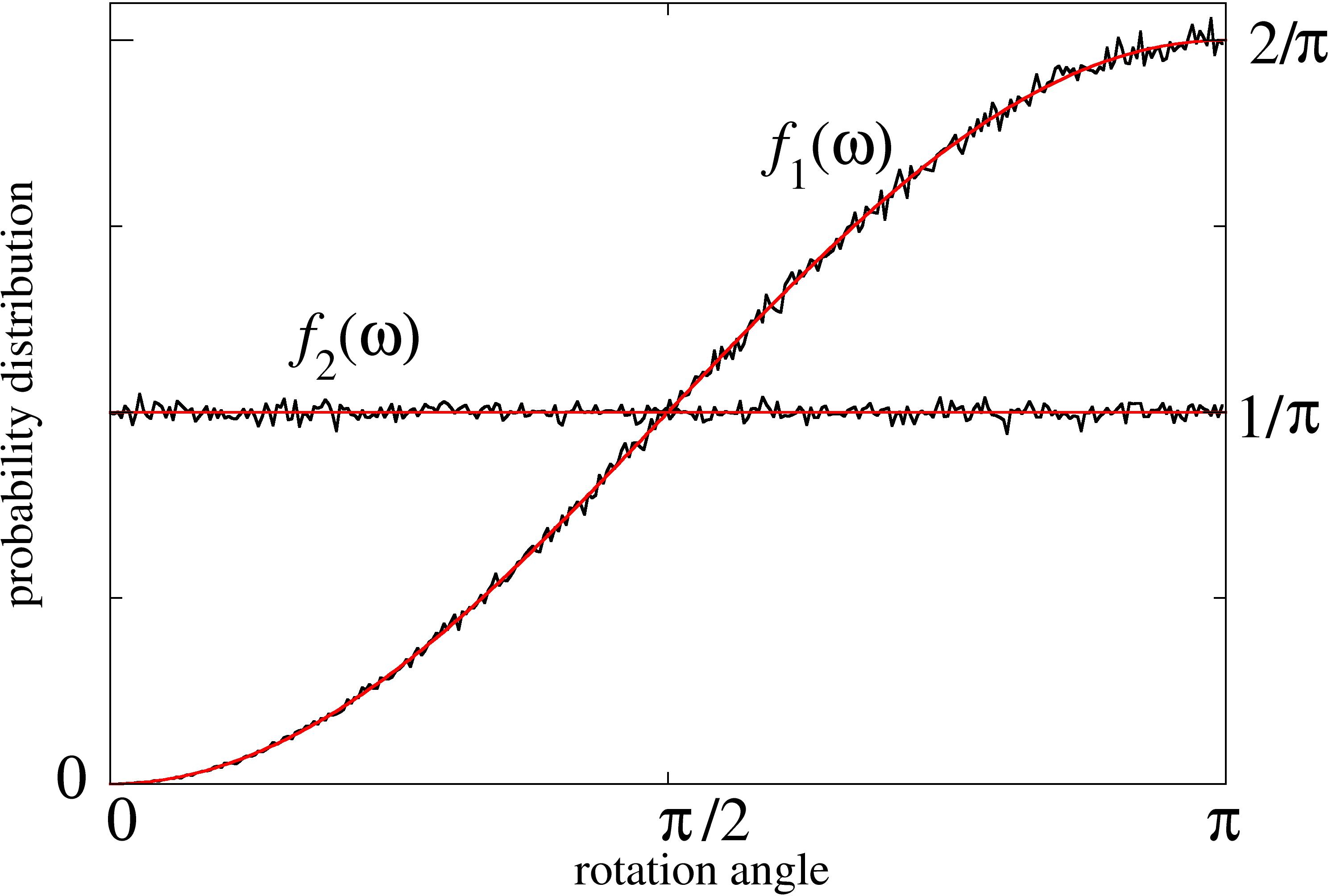}
  \caption{The theoretical Haar measure $f_1(\omega) =
    (1-\cos{\omega})/\pi$ (\eref{eq:f1}) overlayed with a histogram of
    the rotation angle of
    \num{e6} randomly drawn matrices. 
   Similarly, the Haar measure $f_2(\omega) = 1/\pi$
   (\eref{eq:f2}) and the histogram of the rotation angle for \num{e6}
   squared random rotations. The fluctuations are Gaussian (not shown). }
\label{fig:nonflat}
\end{figure}

We are interested in how ``far'' a random walk wanders from the
origin, specifically in the probability  density $f_1(\omega)$ that this walk ends in a shell
of radius $\abs{\r}=\omega$ in \fref{fig:sample1}. Therefore, we
integrate $\dd{\mu}$ over all directions $\n$, and find (\fref{fig:nonflat})
\begin{equation}
  f_1(\omega) \dd {\omega} = \int_{\n \in S^2}\dd \mu = \ltfrac{1}{\pi}\left(1-\cos\omega \right) \dd {\omega} ~.
  \label{eq:f1}
\end{equation}
Random rotations by small angles are therefore relatively rare since the distribution vanishes as $f_1(\omega) \sim 1-\cos{\omega} \sim \omega^2$.

\nins{Powers of random rotations distribute uniformly}
 It is perhaps surprising, but easy to see, that when one 
doubles a random rotation, there is much more weight for small rotation angles as their distribution becomes constant $f_2(\omega)=1/\pi$ (\fref{fig:nonflat}):
Repeating a given random rotation by an angle $\omega_1$ twice implies that the rotation angle is doubled and ``folded'' back into the $[0,\pi]$ domain as
$\omega = \min\{2\omega_1,-2\omega_1 + 2\pi \}$, where we used the antipodal symmetry $\R(\n,\omega) = \R(-\n,-\omega+2\pi)$). Summing over the folds, one obtains the probability density that after two rotations the total angle is $\omega$,
\begin{equa}
  f_2(\omega)&= \ltfrac{1}{2\pi}
  \left(1-\cos{\ltfrac{\omega}{2}}\right)+\ltfrac{1}{2\pi}\left(1-\cos{\ltfrac{2\pi-\omega}{2}}\right) 
  = \ltfrac{1}{\pi}~.
  \label{eq:f2}
\end{equa}

Likewise, we show that for any $m\ge2$ repeats, the distribution of the overall rotation angle remains invariant $f_m(\omega)= 1/\pi$. Using the two symmetries, $\R(\n,\omega) = \R(-\n,2\pi - \omega)$ and $\R(\n,\omega) = \R(\n,\omega \bmod{2\pi})$, one folds the multiplied angle $m \omega_1$ onto $[0,\pi]$, to find the overall angle $\omega_m = \min\{\pm m\omega_1\bmod{2\pi}\}$.
Summing over the $m$ folds, we obtain
\begin{equa}
  f_m(\omega) &= \sum_{j = 0}^{m-1}
   \ltfrac{1}{m\pi}\left(1-\cos{\ltfrac{\omega + 2 j\pi}{m}} \right)\\    
                &= \ltfrac{1}{\pi}-\ltfrac{1}{m\pi}\Re \bigg( e^{i\frac{\omega}{m}}\mathlarger{\sum}_{j=0}^{m-1}e^{i\frac{2\pi j}{m}} \bigg)=\ltfrac{1}{\pi}~, 
\end{equa}
where the sum over the $m$-th roots of unity vanishes by symmetry for any $m \ge 2$.
This calculation can be seen as a special case of
decomposing a Lie group into a sequence of subgroups~\cite{Diaconis_Shahshahani_1987,Rains_2003}. 
Thus, while the distribution of random rotations $\R(\n,\omega)$
vanishes close to $\One$, at $\omega =0$, it is enough to take any
integral power $\R^m$ ($m\ge 2$) to make the distribution exactly
uniform, with many rotations close to $\One$.

This completes the probabilistic argument: It is practically impossible to find a scaling factor $\lambda$ that brings the stretched walk back to the origin because random rotations with small angles are exceedingly rare, $f_1(\omega = 0) = 0$. For a similar reason, it is almost always possible to return a double, stretched walk to the origin because random \SI{180}{\degree}-rotations are quite common; there is a whole 2D sub-manifold of those, $f_1(\omega = \pi) = 2/\pi$. And this argument generalizes to any $m \ge 2$: the solutions $\lambda_\ast$ to the problem $[\Wlam]^m = \One$ are the factors $\lambda_\ast$ for which $\W(\lambda_\ast)$ are rotations by any of the angles $\theta_{\mathrm{single}} = 2 j \pi/m$, for $j = 1,\ldots,\lfloor m/2 \rfloor$.
Due to the double covering of $\SO(3)$ by $\SU(2)$, the same argument applies to randomly distributed unitary matrices $\U \in \SU(2)$.
\footnote{The 2:1 homomorphism between $\SO(3)$ and $\SU(2)$ is
$\pm \U(\n,\half\omega) = \pm \exp \textstyle{[} \frac{i}{2}\omega (\n \vdot \vsig) \textstyle{]} = \pm  \textstyle{[} \cos{\half \omega}\cdot \One + i \sin{\half\omega}(\n\vdot\vsig)\textstyle{]} \mapsto \R(\n,\omega ) = \exp \textstyle{[} \omega (\n \vdot \pmb{L}) \textstyle{]} 
$, where $\vsig = (\sig_x,\sig_y,\sig_z)$ are Pauli's matrices.
The mapping implies that the angle distribution in $\SU(2)$ is the same as in $\SO(3)$ (\eref{eq:f1}), only for half-angles~\cite{rummler2002}.
In particular, $\pm \U(\n,0) = \mp \U(\n,\pi) = \pm \One$, which map to the identity rotation $\R(\n,0)  = \R(\n,2\pi) = \One$, are rare. But $\U(\n,\pi/2)$ are abundant and map to the identity when squared, $[\U(\n,\pi/2)]^2 = \U(\n,\pi) = -\One \mapsto \R(\n,2\pi) = \One$}

\begin{figure}[htb!]
\centering\includegraphics[width=1\columnwidth]{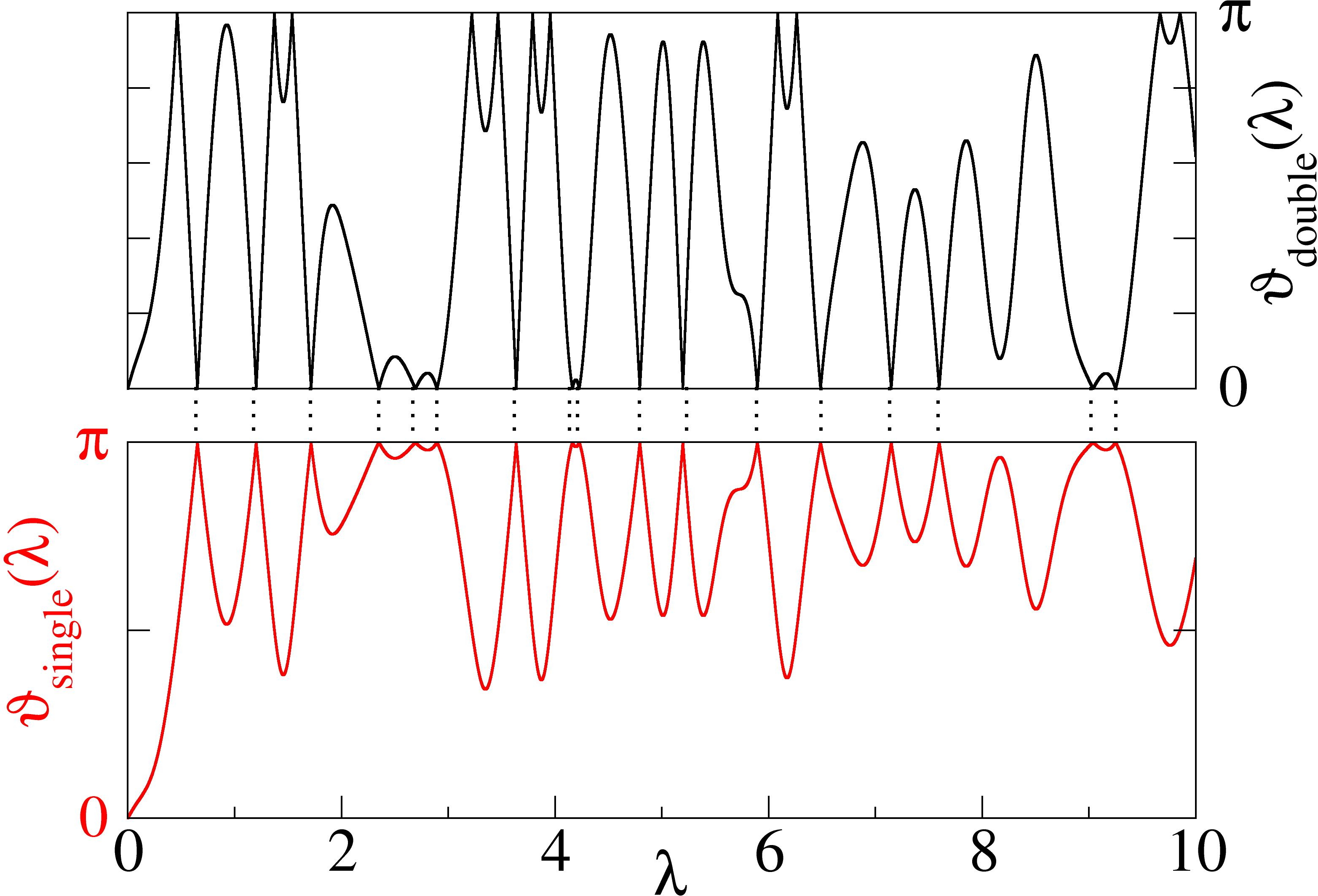}
\caption{The dependence of the overall rotation angle on the scaling factor $\thetasing(\lambda)$ for a random stretched walk $\Wlam = \prod_{i=1}^\N (\R_i)^\lambda$ (bottom, red line) and the rotation angle for the twice-repeated walk $[\Wlam]^2$ (top, black line). The walk is composed of $\N = 40$ randomly drawn rotations with the Haar measure. The twice-repeated stretched walk returns to the origin in several points, $\thetadoub(\lambda_\ast) = 0$, where the single repeat angle is $\thetasing(\lambda_\ast) = \pi$ 
(connected by dots). The mathematical question consists in showing that for each given series of rotations $\W$ there exists at least \emph{one} $\lambda_\ast$ where $\thetasing(\lambda_\ast)=\pi.$}\label{fig:stretching}
\end{figure}

\nins{The trigonometric Diophantine problem} While the probabilistic argument above shows how the chance of returning close to the identity increases when the walk is doubled and stretched, we will now argue that by rescaling with $\lambda$, one can return \emph{precisely} to the identity, as demonstrated in \fref{fig:stretching}.
To this end, consider a product of rotations (a ``walk'') in $\SO(3)$, $\W(1) = \R_\N \cdot \ldots\cdot
\R_1$. We ask under which conditions $\Wlam 
\equiv \R_\N^ \lambda \cdot \ldots\cdot \R_1^ \lambda$ can become the identity. 

To study this question, we start from a given random walk, $\W =\prod_{j=1}^\N \R_{j}$, where the $\R_{j}$ are drawn with the Haar measure and trace how the ``stretched'' walk scales, $\W \to \Wlam =  \prod_{j=1}^\N (\R_{j})^\lambda$, as we vary $\lambda$ over a large range $[0,\lambda_{\rm max}]$. 
We extract the overall rotation angle from the trace or the logarithm, as $\thetasing(\lambda) = \norm{\log{\Wlam}} = \arccos[\half(\Tr\Wlam-1)]$ (\fref{fig:stretching}~bottom). 
Apart from the trivial limit, ${\thetasing(\lambda)} \to 0$ when $\lambda \to 0$, we see that $\Wlam$ does not return to the origin, so nontrivial solutions for $\Wlam = \One$ are lacking.
However, in the same range of $\lambda$, the doubled random walk
$[\Wlam]^2$ returns to the origin several times
(\fref{fig:stretching}~top), $\thetadoub(\lambda_\ast) =
0$, exactly for those $\lambda_\ast$ where $\thetasing(\lambda_\ast) = \pi$.

Our main result is proven in two parts: First, we show that the set of
rotations and $\lambda$ factors for which $\Wlam$ itself reaches the
identity has a small measure (Lemma below). Second, we will show that, in contrast,
there are ``many'' $\lambda>0 $ for which $[\Wlam]^2 = \One$ (Theorem below).
Throughout the proofs, we use the following notation: We consider rotations $\R_k$ by an angle $\omega_k$ around an axis $\n_k$.
One can safely assume that axes of consecutive rotations are not collinear, $\abs{\n_k\vdot\n_{k+1}} < 1$,
otherwise, they can be trivially composed into one rotation. Finally, we define the partial product of the $\lambda$-powers of the first $j$ rotations, $\w(\lambda)_j \equiv \R_j^\lambda \cdot \R_{j-1}^\lambda \cdot\ldots\cdot \R_1^\lambda $, whose \emph{total} rotation angle is $\phi(\lambda)_j$, around an axis $\a(\lambda)_j$. \\

\noindent\textbf{Lemma} (Single walks returning to $\One$ are rare):
\emph{Given a walk $\W = \prod_{j=1}^\N{\R_j}$, there are three trivial
cases: (i) $\W(\lambda=0) = \One$, no rotation; (ii) $\W$ itself is the identity, $\W(\lambda =1)=\One$; (iii) $\W$ is a
single rotation, $\W(1) = \R(\n,\omega)$, so $\W(2\pi/\omega) = \One$. In all other non-trivial cases, the identity
can only be reached if all rotation axes are collinear, or if all
rotation angles are commensurate, which are cases of negligible
measure.}  
\begin{proof}
We recursively decompose the rotation product using the Rodrigues
formulae, which specify the overall angle and axis of
rotation when two rotations are applied successively (Eqs.~(19,20) in~\cite{altmann1989}, originally published in~\cite{rodrigues1840}). 
The first decomposition step, $\Wlam =
\R_\N^\lambda \cdot \w(\lambda)_{\N-1}$, yields:
\begin{equa}[eq:rodrigues]
   \cos{\ltfrac{\phi(\lambda)_\N}{2}} = \cos{\ltfrac{\lambda \omega_\N}{2}}&\cos{\ltfrac{\phi(\lambda)_{\N-1}}{2}} \\
  - \sin{\ltfrac{\lambda\omega_{\N}}{2}} &\sin{\ltfrac{\phi(\lambda)_{\N-1}}{2}} 
     \left[\n_\N\cdot\a(\lambda)_{\N-1} \right]  =\\
   x(\lambda)_\N \cos{\ltfrac{\lambda \omega_\N +  \phi(\lambda)_{\N-1}}{2}} &
    +[1-x(\lambda)_\N] \cos{\ltfrac{\lambda \omega_\N -\phi(\lambda)_{\N-1}}{2}}~,
  \end{equa}
where $x(\lambda)_k \equiv\half[1+\n_k\cdot \a(\lambda)_{k-1}]$ (thus, $0 \le x_k(\lambda) \le 1$). 
Rodrigues formulae are essentially the multiplication rule of $\SU(2)$ matrices, and the abundant halves originate from mapping rotations to unitary matrices with half-angles~\cite{rummler2002,Note1}. 

Now, solving $\Wlam = \One$ implies that $\phi(\lambda)_\N = 0$ , and therefore $\cos\half\phi(\lambda)_\N = 1$
in \eref{eq:rodrigues}. This condition is fulfilled in two cases:
(i) $x(\lambda)_\N\in \{0,1\}$, which means that the rotations are collinear, $\abs{\n_\N\cdot \a(\lambda)_{\N-1}}=1$ or (ii) both cosines equal one, $\cos[\tfrac{1}{2}(\lambda \omega _\N\pm\phi(\lambda)_{\N-1})] =1$.
In the first case, we observe that the condition $x(\lambda)_\N\in\{0,1\}$ is of
measure 0 in the choice of $\a(\lambda)_{\N-1}$ once $\n_\N$ is given; this condition is of codimension 2. 

The second case is also rare because it implies
$[\lambda \omega _\N+\phi(\lambda)_{\N-1}]/2=2\alpha _\N\pi$, and  
$[\lambda \omega _\N-\phi(\lambda)_{\N-1}]/2=2\beta _\N\pi$,
with integer $\alpha _\N,\beta _\N$. It follows that
$\lambda \omega _\N=2(\alpha _\N+\beta _\N)\pi$, and 
$\phi(\lambda)_{\N-1} =2(\alpha _\N-\beta _\N)\pi$, 
and therefore $\cos{\half\phi(\lambda)_{\N-1}}=1$, 
so we can recursively decompose $\cos{\half\phi(\lambda)_{\N-1}}$, exactly as we did in \eref{eq:rodrigues}.
Through this induction, we find 
\begin{equa}
  \lambda \omega _k= 2(\alpha_k+\beta_k)\pi,\text{ for } k=1,\dots, \N~.
\end{equa}
This means that $\Wlam  \ne \One$ for any $\lambda\not\in\{0,1\}$, unless all the $\omega _k$,
$k=1,\dots,\N$, are commensurate (all ratios $\omega_j/\omega_k$ are rational numbers). This completes the proof.
\end{proof}

We now show that there actually exist $\lambda>0 $ for which $[\Wlam]^2
= \One$, using a technique of Conway-Jones~\cite{ConwayJonesICM}, and a
Diophantine argument by Minkowski~\cite{Minkowski2002}[Ch. 2].
(Note that we are not dealing with a small-divisor problem \cite{Yoccoz1992}.)\\
  \noindent\textbf{Theorem}~(Double walks returning to $\One$ are abundant):
  \emph{Let $\{\omega _j>0, j=1,\dots, \N\}$ be the set of angles of rotation of $\N$ rotations
  $\R_j, j=1,\dots,\N$. Then, when the set $\{\omega _j\}$ does not consist
  only of pairs of equal angles, there is a $\lambda>0 $ for
  which $[\W(\lambda )]^2=[\R_\N^\lambda \cdot \R_{\N-1}^\lambda \cdot\ldots\cdot
  \R_1^\lambda]^2 = \One$.}

\begin{proof}
We consider first the walk $\Wlam$ and show, by a continuity argument, that the overall rotation angle $\phi_
{\N}(\lambda)$ can be made to be equal to $\pi$ for some $\lambda =\lambda_*
$, and therefore $\W(\lambda_*)^2$ will have a rotation angle of $0$,
which means it is the identity. 

To find $\lambda_*$, we return to the Rodrigues formulae, which will also eventually explain the condition forbidding pairs of equal angles. 
The formulae for the rotation product $\R(\a_2,\phi_2)=\R(\n_2,\omega_2 )\cdot \R(\n_1,\omega_1)$ are
(Eqs.~19 (\eref{eq:rodrigues}) and~20 in~\cite{altmann1989}):
\begin{equa}[eq:rod2]
  \cos{\ltfrac{\phi_2}{2}}&= \cos{\ltfrac{\omega_2}{2}}\cos{\ltfrac{\omega_1}{2}}-
  \sin{\ltfrac{\omega_2}{2}}\sin{\ltfrac{\omega_1}{2}}(\n_2\vdot\n_1)~,\\
   \sin{\ltfrac{\phi_2}{2}}&=\sin{\ltfrac{\omega_2}{2}}\cos{\ltfrac{\omega_1}{2}} (\n_2\vdot\a_2)\\
  + &\cos{\ltfrac{\omega_2}{2}}\sin{\ltfrac{\omega_1}{2}}(\n_1\vdot\a_2)
  +\sin{\ltfrac{\omega_2}{2}}\sin{\ltfrac{\omega_1}{2}} (\n_2\cross\n_1)\vdot\a_2~. 
\end{equa}
Applying \eref{eq:rod2} recursively for arbitrary finite products of $\N$ rotations, we obtain
\begin{equ}\label{eq:sum}
F(\lambda )\equiv\cos{\ltfrac{\phi(\lambda)_\N}{2}}=\prod_{j=1}^{\N }{\cos{\ltfrac{\lambda\omega_j}{2}}}  +X~,
\end{equ}
where each term in the remainder $X$ is a product of sines and cosines with at least one sine. Furthermore, the coefficients of all these terms are bounded by 1 in absolute value, because they are various dot and cross products of unit vectors 
\footnote{Rodrigues formula (\eref{eq:rod2}) and the resulting \eref{eq:sum} can be derived by expanding products of the equivalent $\SU(2)$ matrices, $ \R(\n,\omega) \to \U(\n,\half\omega)= \cos{\half\omega} \cdot \One+ i \sin{\half\omega} (\n\cdot \vsig)$, using  Pauli's identity, $(\b\vdot \vsig)(\c\vdot \vsig) =  (\b\vdot \c) \cdot \One + i(\b \cross \c )\vdot \vsig$. The products of $\cos{\half\lambda\omega_j}$ and $\sin{\half\lambda\omega_j}$ in \eref{eq:sum} can be transformed into cosines and sines of sums of $\pm{\half\lambda\omega_j}$. So, $F(\lambda)$ is a sum of periodic functions, explaining the oscillations in \fref{fig:stretching} }.

We want to show that there is a $\lambda_*$ for which $F(\lambda_*) =
\cos{\half\phi(\lambda_*)_\N} = 0$, and therefore $\phi(\lambda_*)_\N =
\pi \bmod 2\pi$, so that $\W(\lambda_*)$ is a \SI{180}{\degree}-rotation, which when repeated twice gives the identity. 
We use the continuity of $F(\lambda)$: At the limit $\lambda = 0$, the
overall angle vanishes, $\phi(\lambda)_\N = 0$, and $F(0) = 1$. It is
therefore enough to show that $F(\lambda_{-}) < 0$ somewhere to prove
that there is $0 < \lambda_* < \lambda_{-}$ for which
$F(\lambda_*)=0$.
The idea is to construct a $\lambda_{-}$ for which all $\cos{\half\lambda\omega_j}$ are \emph{simultaneously} close to $-1$ and all $\sin{\half\lambda\omega_j}$ are close to $0$. This is for an odd number of rotations $\N$; for an even $\N$, we require that one cosine is close to $+1$ and all others close to $-1$. This is the reason for the ``no pairing'' condition in the theorem, because, otherwise, we cannot single out one such positive $\cos{\half\lambda\omega_j}$.

Showing that $F(\lambda)$ (\eref{eq:sum}) can become negative is really a
(trigonometric) Diophantine problem~\cite{ConwayJonesICM}:
We would like $\prod_{j=1}^\N\cos{\half\lambda\omega_j} +X$ to be close to $-1$.
So, we want all $\half\lambda \omega_j$ to be close to an odd multiple of $\pi$ (\ie $n_j\pi = (2 k_j+1)\pi$ for some $k_j\in\integer$) because then the cosines are all close to $-1$ and so is their product in (\eref{eq:sum}), whereas the other terms in $X$ are vanishingly small because they contain sines. For an even number of cosines, we ask that a single $\half\lambda \omega_j$ is close to an even multiple $n_j\pi = 2 k_j\pi$ (see below).
We define $n_0 \equiv \tfrac{\lambda }{2\pi}$ and rewrite this requirement as $\omega_j n_0\simeq n_j$.
And thus, we seek 
\begin{equa}[eq:ineq]
  \abs{\omega_j n_0-n_j} <\frac{4\epsilon}{\L}~, \text{ for all } j=1,\dots,\N  ~,
\end{equa}
where $\L$ is the finite number of terms in $X$. If we apply (\ref{eq:ineq}) to the product of the $\cos(\half\lambda
\omega_j)=\cos(\pi n_0 \omega_j)$, with $n_j=2k_j+1$, we see that each cosine is close to $-1+8\pi^2 \epsilon ^2/\L^2$  while each $|\sin(\half\lambda \omega _j)|\le 4\pi\epsilon/\L $. Thus, the product in
\eref{eq:sum} is smaller than $(-1+8\pi^2\epsilon^2/\L^2)^\N$ and the remainder
 is bounded, $\abs{X} \le 4\pi \epsilon$. So, if $\N$ is odd, $F(\lambda)$ is order of $\epsilon$ close to $-1$.

To solve the Diophantine problem, (\ref{eq:ineq}), we use Minkowski's theorem~\cite{minkowski1910,Minkowski2002}:
\emph{Let $\Lambda \subset \real^M$ be a lattice with a co-volume $\cov(\Lambda)$, and let $S \subset \real^M$ be a convex set symmetric with respect to the origin 0, having a volume $V(S)$.
Then, if  $V(S) > 2^M\cov(\Lambda)$, the set $S$ contains a non-zero lattice point.}

To apply this to our case, we consider the lattice $\Lambda \in \real^{1+\N}$ whose points are $(\N+1)$-tuples $(n_0,n_1,\dots,n_\N) \in \integer^{1+\N}$. To ensure that the cosine product in \eref{eq:sum} is close to $-1$,  all $\N$ coordinates $n_1,\dots,n_\N$ are odd integers,  if $\N$ is odd, and if $\N$ is even, $\N-1$ coordinates are odd and one is even, say $n_\N$; the coordinate $n_0$ is any integer. Thus, the distance between neighboring lattice points is $2$ along the directions $n_1,\dots,n_\N$ and $1$ along the $n_0$ direction.  The unit cell of $\Lambda$ is therefore a hyperrectangle of volume $\cov(\Lambda) = 2^{\N}$ ($\Lambda$'s co-volume). 
Then, for a small given $\epsilon >0$, we define the convex set 
\begin{equa}
  S&=\bigl\{ (n_0,n_1,\dots,n_\N)\in \real^{1+\N}~:~\\& | n_0|\le
 \frac{\L^\N}{\epsilon^\N} , \quad|\omega_j n_0 -n_j| \le \frac{4\epsilon}{\L},~ j=1,\dots,\N\bigr\}~.
\end{equa}
This set is symmetric w.r.t.~the origin ($S = - S$) and has a volume $V(S) = 2^{1+3\N}  > 2^{1+\N} \cov(\Lambda) = 2^{1+2\N}$. 
Hence, Minkowski's theorem applies and there are non-zero lattice points in this set, $(n_0,n_1,\dots,n_\N) \in S \cap \{ \Lambda \setminus 0 \}$, satisfying the inequalities (\ref{eq:ineq}). This completes the proof. 
\end{proof}
\noindent The convex set $S$ is an extremely long (${\sim}(\L/\epsilon)^\N$) and thin (${\sim}\epsilon/\L$) needle-like volume, and therefore the non-zero lattice points may be very far from the origin, $\abs{n_j/\omega_j} \simeq \abs{n_0} \lesssim (\L/\epsilon)^\N$, so the scaling factor, $\lambda = 2 \pi n_0$, is large.

Note that the proof implies that solutions $\lambda_m$ of $[\W(\lambda_m)]^m = \One$ also exist for any $m > 2$. It follows from the continuity of $F(\lambda)$ that there is $0<\lambda_m<\lambda_{\ast}$, for which $F(\lambda_m) = \cos(\pi/m)$, since $1 = F(0) > F(\lambda_m) = \cos(\pi/m) > F(\lambda_{\ast}) =0$. Thus, the rotation angle of $\W(\lambda_m)$ is $2\pi/m$, so $[\W(\lambda_m)]^m = \One$.

\nins{Conclusion}
By combining the Rodrigues formulae with the Minkowski theorem, we found that almost any sequence of rotations in $\SO(3)$ or $\SU(2)$ (equivalently, any arbitrary field $\B(t)$) may return a system to its original state if this series is scaled and repeated more than once. Finding such a scaling amounts to solving a trigonometric Diophantine equation, and the solution applies to any physical system governed by $\SO(3)$ or $\SU(2)$, such as magnetic spins or qubits.

One immediately wonders whether this scenario generalizes to $\SO(d)$ with a dimension $d>3$. 
A rotation in $\R \in \SO(d)$ is specified by  $\lfloor \frac{d}{2} \rfloor$ rotation angles, $\omega_\alpha$, and their corresponding invariant planes~\cite{gallier2002}. The depletion of random matrices near $\One$ becomes even more severe as the dimension $d$ increases, as evident from the diminishing Haar measure around $\omega_\alpha=0$~\cite{rummler2002,meckes2019},  
which scales as
$\dd\mu \sim  \prod_{\alpha<\beta}{(\omega_\alpha^2-\omega_\beta^2 )^2}$ (for~even~$d$; for odd $d$, this is multiplied by  $\prod_{\alpha }{\omega_\alpha^2}$).

Indeed, numerical experiments similar to \fref{fig:stretching} show that finding a scaling factor for which $\Wlam=\One$ is hopeless. However, for $d >3$, repeating the walk $m\ge2$ times does not suffice to find $[\Wlam]^m=\One$, because now all $ \lfloor \frac{d}{2} \rfloor > 1$ rotation angles $\omega_\alpha$ should simultaneously cross some multiple of $2\pi/m$. 
Put differently, the identity roots lie on a manifold of codimension $ \lfloor \frac{d}{2} \rfloor$ (and dimension $\binom{d}{2}-  \lfloor \frac{d}{2} \rfloor$). One may therefore conjecture that crossing this manifold generically requires scaling in $\lfloor \frac{d}{2} \rfloor$ independent directions, each governed by its own $\lambda$.
\begin{acknowledgments}
JPE was supported by Fonds National Suisse Swissmap, and TT by the National Research Foundation of Korea Grants NRF-RS-2025-00573354. The authors thank Yaroslav Sobolev, Mo Bahrami, and an anonymous referee for helpful comments. 
\end{acknowledgments}

\bibliography{Ref}

\end{document}